\documentclass[pre,onecolumn,showpacs,preprintnumbers,amsmath,amssymb]{revtex4}
\usepackage{graphics,bm,rotating}

\begin{document}

\title{
Spectroscopic data for the LiH molecule from pseudopotential quantum Monte 
Carlo calculations
}
\author{J. R. Trail}\email{jrt32@cam.ac.uk}
\author{R. J. Needs}
\affiliation{TCM Group, Cavendish Laboratory, University of Cambridge,
J J Thomson Avenue, Cambridge, CB3 0HE, UK}

\date{April, 2008}

\begin{abstract}
Quantum Monte Carlo and quantum chemistry techniques are used to
investigate pseudopotential models of the lithium hydride (LiH)
molecule.  Interatomic potentials are calculated and tested by
comparing with the experimental spectroscopic constants and well
depth.  Two recently-developed pseudopotentials are tested, and the
effects of introducing a Li core polarization potential are
investigated.  The calculations are sufficiently accurate to isolate
the errors from the pseudopotentials and core polarization potential.
Core-valence correlation and core relaxation are found to be important
in determining the interatomic potential.
\end{abstract}


%02.70.Ss Quantum Monte-Carlo methods
%31.15.vn Electron correlation calculations for diatomic systems
%71.15.Dx Computational methodology (Brillouin zone sampling, 
%         iterative diagonalization, pseudopotential construction)
%33.20.Vg Vibration-rotation analysis

%keywords:
%Lithium Hydride, pseudopotential, quantum monte-carlo, vibration-rotation

\maketitle

\section{introduction}

Pseudopotentials are often generated using data from mean-field
theories such as density-functional theory and Hartree-Fock (HF)
theory, but they are often used in more accurate many-body approaches,
such as Multi-Configuration Self-Consistent Field (MCSCF) and quantum
Monte Carlo (QMC) calculations.  In this paper we apply such methods
to a simple system, the LiH molecule with pseudopotentials
representing the Li$^+$ and H$^+$ ions, for which we can obtain very
accurate ground state energies.  Our main aim is to investigate the
accuracy of pseudopotentials for this system, including the
performance of a Li core-polarization potential (CPP) which introduces
some of the effects of core-valence correlation and core relaxation.
The accuracy of the LiH interatomic potentials is measured by
calculating the spectroscopic constants of the diatomic molecule and
the well-depth, which can readily be compared with experiment.  Note
that our goal is not to achieve the most accurate \emph{ab initio}
spectroscopic constants - if it were we would solve the all-electron
system directly using QMC, or one of the range of accurate methods
tractable for a four electron system - but to separate the
deficiencies of a model valence Hamiltonian from those of the methods
available for its solution.

Our most accurate results are obtained with the variational quantum
Monte Carlo (VMC) method and the more sophisticated diffusion quantum
Monte Carlo (DMC) method.\cite{foulkes_2001} Although the scaling of
the computational cost of these calculations with particle number $N$
is reasonable ($\sim$$N^3$\cite{foulkes_2001}), the cost increases
rapidly with $Z$ ($\sim$$Z^{5.5}$\cite{ma05,ceperley86}). It is
therefore normal to use pseudopotentials for heavier atoms.  It is
highly advantageous to use pseudopotentials which are smooth at the
origin in VMC and DMC calculations, but most of the pseudopotentials
found in the quantum chemistry literature diverge at the origin.  We
have tested two sets of recently published
pseudopotentials\cite{trail05a,trail05b,burkatzki07} for the Li$^+$
and H$^+$ ions, which are smooth at the origin and have been designed
for use in QMC calculations.  One of these is a ``shape-consistent''
pseudopotential\cite{trail05a,trail05b} generated from the
Hartree-Fock atomic ground state while the other is an
``energy-consistent'' pseudopotential\cite{burkatzki07} generated from
the ground and excited state energies of the Hartree-Fock atom.

The pseudopotential model of the LiH diatomic molecule contains two
valence electrons of opposite spin.  The ground state wave function is
therefore nodeless, so that no error arises from the ``fixed-node
approximation'' used to enforce the fermion antisymmetry in DMC.
However, the DMC energy is not exact, even in principle, because the
non-local pseudopotential is treated approximately, although we
demonstrate that the error must be small in this case.  Our
calculations are therefore exacting tests of the accuracy of the
pseudopotentials and CPP.

Throughout we consider the error of the LiH diatomic molecule with
both Li and H described by pseudopotentials. No attempt is made to
separate the performance of each pseudopotential, but atomic
calculations suggest that errors arising from deficiencies of the H
pseudopotential will be an order of magnitude smaller than those from
the Li pseudopotential.

The quality of a diatomic potential can be characterised by the
spectroscopic constants, or Dunham coefficients,\cite{dunham32} which
can be determined experimentally from the rotation/vibration spectrum.
Calculations which include an accurate description of electron
correlation are computationally expensive and so they are normally
performed at a small number of geometries and the interatomic
potential is obtained by fitting to a suitable functional form.
Fitting to a small number of energies introduces an error which is
further exacerbated if the energies have a statistical uncertainty, as
in QMC estimates.  We explore this problem carefully and ensure that
such errors are small.  In what follows only the $X^1 \Sigma^+$
(ground) state of the $^7$Li$^1$H isotopologue is considered.

\section{spectroscopic constants}

A Born-Oppenheimer decoupling of the electron/nucleus coordinates of a
diatomic molecule leads to the energy levels
\begin{equation}
E_{vK}=\sum_{ij} Y_{ij} (v+1/2)^i K^j(K+1)^j,
\label{eq:1}
\end{equation}
where $v$ and $K$ are the vibrational and rotational quantum numbers,
and the spectroscopic constants, $Y_{ij}$, depend on the underlying
interatomic potential.\cite{dunham32} Not all of these coefficients
may be extracted directly from experimental data - $Y_{00}$ cannot
since it does not influence the spacing of levels, although it is
required to define the zero point energy $E_{\rm ZP}=E_{00}$.  We
follow Dunham and express $Y_{00}$ as
\begin{equation}
Y_{00} = \frac{Y_{01}}{4} - 
         \frac{Y_{11}Y_{10}}{12Y_{01}} +
         \frac{Y_{11}^2Y_{10}^2}{144Y_{01}^3} +
         \frac{Y_{20}}{4} + 
         \ldots
\label{eq:2}
\end{equation}
for experimental data, while $Y_{00}$ is available directly from
\emph{ab initio} data.

The zero point energy is given by
\begin{equation}
E_{\rm ZP} = Y_{00} + \frac{1}{2}Y_{10} + \frac{1}{4}Y_{20} +
\frac{1}{8}Y_{30} + \frac{1}{16}Y_{40} + \ldots,
\end{equation}
and the `harmonic equilibrium separation', $R_e$, is defined as
\begin{equation}
R_e^2 = \frac{1}{2 \mu} \frac{1}{Y_{01}},
\label{eq:3}
\end{equation}
where $\mu$ is the reduced nuclear mass of the system, and all
quantities are in atomic units.  Finally, we note that the
spectroscopic constants are normally defined as
\begin{eqnarray}
w_e      &=&  Y_{10} \nonumber \\
w_ex     &=& -Y_{20} \nonumber \\
w_ey     &=&  Y_{30} \nonumber \\
B_e      &=&  Y_{01} \nonumber \\
\alpha_e &=& -Y_{11} ,
\end{eqnarray}
although we use the $Y_{ij}$ notation throughout.

To obtain estimates of $\{Y_{ij}\}$ from a set of total energy values
requires some further analysis.  We fit a highly flexible form for the
interatomic potential to a finite number of total energies evaluated
at different geometries.  We use the ``modified Lennard-Jones
oscillator''\cite{coxon04,modpot_extra} potential,
\begin{equation}
U(R) = U_{\infty}-D_e+D_e \left[ 1-\left( \frac{R_0}{R} \right)^6 
 e^{-\phi(z) z} \right],
\label{eq:4}
\end{equation}
where $U(R)$ is the total energy for interatomic spacing $R$, $R_0$ is
the position of the minimum in the interaction potential, $D_e$ is the
well depth, $U_{\infty}$ is the large $R$ limit of $U(R)$, and
$\phi(z)$ is defined by
\begin{equation}
\phi(z) = \sum_{m=0}^{M} a_m z^m
\end{equation}
for some choice of $M$, where
\begin{equation}
z = 2\frac{(R-R_0)}{(R+R_0)} .
\end{equation}
A non-linear least-squares fit to $U(R)$ (with $U_{\infty}$ provided
by isolated atom calculations using the same method as the diatomic
calculations) provides the parameters $(\{a_m\},D_e,R_0)$. Note that
the bond dissociation energy, $D_0$, is related to the well depth and
the zero point energy by $D_e=D_0+E_{\rm ZP}$.  The derivatives of
$U(R)$ at $R_0$ may be evaluated and used together with Dunham's
formulae\cite{dunham32} to obtain the spectroscopic constants and the
values of $D_e$ and $R_e$.
\footnote{Note that Dunham's formulae are approximate. We could
explicitly solve the Schr\"odinger equation for this potential and
then perform a least-squares fit of the eigenenergies to
Eq.~(\ref{eq:1}), giving the spectroscopic constants. Although this
might be slightly more accurate, tests indicate that it makes no
significant difference to the accuracy achieved in LiH.}

\section{errors in estimating the spectroscopic constants}

The above procedure gives the spectroscopic constants from the total
energies at a finite number of geometries.  To apply the procedure we
choose a set of sample geometries and the number of free parameters in
Eq.~(\ref{eq:4}) ($M+3$).  For the QMC calculations it is also
necessary to choose a target statistical error bar for the QMC
energies.  Generally speaking, for a fixed statistical error in each
energy point, using more data points reduces both the statistical
error and systematic bias in the spectroscopic constants, while using
data points covering a smaller range of $R$ increases the statistical
error and reduces the bias.  Using fewer parameters in
Eq.~(\ref{eq:1}) reduces the statistical error in the estimated
spectroscopic constants but increases the bias.  An example of
choosing suitable geometries so as to obtain acceptably small
statistical errors and biases in a fit to QMC energies is described in
Ref.~\cite{maezono07}, where the equation of state of diamond is
estimated.

To study these effects we need a reasonably accurate model of the
interatomic potential, for which we use the one constructed recently
for LiH by Coxon and Dickinson,\cite{coxon04} which accurately
reproduces a wide range of experimental spectroscopic data.  We will
refer to this as the CD potential. We take energies from the CD
potential at a finite number of geometries and determine the resulting
errors in the spectroscopic constants using the derivatives of the
potential at $R_0$ and Dunham's equations.  We then add random noise
to the energies and determine a second set of errors in the
spectroscopic constants by averaging over the noise.

A set of results from this procedure are given in Table~\ref{tab:1}
for $M+3=9$ free parameters and nine geometries characterised by
interatomic distances distributed evenly over $\pm 40\%$ about
$R=1.596$~\AA\ (an estimate of the equilibrium bond length), and a
statistical error in the energies of $10^{-6}$~a.u.  The data in the
second row of Table~\ref{tab:1} differ from the first row by an amount
typical of the variation between experimental estimates, which
illustrates the very high accuracy of the spectroscopic constants
obtained from the CD potential and Dunham's equations.  Comparing the
second and third rows of Table~\ref{tab:1} gives the systematic bias
due to using only nine geometries, while the fourth row gives the
additional random error due to the noise in the energies.  The net
effect on the spectroscopic constants of the bias and random error is
small.  For all eight quantities the random error is larger than the
bias due to sampling at nine geometries, and the only quantities that
differ by more than the statistical error are the values of $D_e$ from
experiment and the model potential.  The statistical errors in the
total energies calculated in this work are in the range $0.5-2.4
\times 10^{-6}$~a.u., and the corresponding errors in the
spectroscopic constants are not significantly larger than those
reported in Table~\ref{tab:1}.

\begin{widetext}
\begin{sidewaystable}
\centering
\begin{tabular}{l|l|lllllll}
Method & $R_e$ & $Y_{10}$ & $-Y_{20}$ & $Y_{30}$ & $Y_{01}$ & $-Y_{11}$ & $E_{\rm ZP}$ & $D_e$ \\ \hline \hline
Exp.\cite{stwalley93}  &
$1.595584  $ &$1405.50936 $ &$23.17938 $ &$0.176365$ &$7.5137510$ &$0.2164606$ &$697.95    $ &$20287.7(3)  $ \\
Model\cite{coxon04} + Dunham &
$1.595594  $ &$1405.511387$ &$23.182699$ &$0.179640$ &$7.513778 $ &$0.216450 $ &$697.953612$ &$20286.000000$ \\
Model\cite{coxon04} + Dunham + finite sampling &
$1.595601  $ &$1405.516456$ &$23.181345$ &$0.173479$ &$7.513712 $ &$0.216316 $ &$697.953791$ &$20285.999963$ \\
Estimated error &
$0.00007   $ &$0.5$         &$0.3$       &$0.1$      &$0.0006$    &$0.001    $ &$0.1$        &$0.2$          \\
\end{tabular}
\caption{Comparison of experimental and calculated values of $R_e$,
$Y_{ij}$, $E_{\rm ZP}$, and $D_e$.  The first row contains the
experimental results of Maki \emph{et al.}\cite{stwalley93} The second
row gives the results obtained from the CD potential\cite{coxon04} and
Dunham's equations.  The third row gives the results obtained by
sampling the CD potential\cite{coxon04} at nine geometries and
re-fitting to the same functional form, and then using Dunham's
equations.  The fourth row gives the standard error in each parameter
obtained from the same procedure as for the data in the third row, but
with added noise corresponding to a standard error in the energies of
$10^{-6}$~a.u.  The energies in the table are in cm$^{-1}$ while $R_e$
is in \AA.}
\label{tab:1}
\end{sidewaystable}
\end{widetext}

\section{description of the methods}

We have performed Hartree-Fock (HF) calculations, post Hartree-Fock
Multi-Configuration Self-Consistent Field (MCSCF) calculations, using
both all-electron (AE) and pseudopotential methods, and
pseudopotential QMC calculations.  We will denote the
``shape-consistent'' pseudopotentials of Trail and
Needs\cite{trail05a,trail05b} by TN and the ``energy-consistent''
pseudopotentials of Burkatzki, Filippi, and Dolg\cite{burkatzki07} by
BFD.  Both of these sets of pseudopotentials were generated using data
from atomic calculations in which electron correlation is neglected.
Both the TN and BFD pseudopotentials contain an approximate
description of relativistic effects, which are small in LiH.
\footnote{For TN, the atomic Dirac-Fock equations are solved at the
outset to provide relativistic pseudopotentials, which are then
reduced to an ``averaged relativistic effective potential'' (AREP).
For BFD the scalar relativistic Wood-Boring equations are solved,
which then provide scalar relativistic pseudopotentials.}

For the HF and MCSCF calculations we used the GAMESS\cite{gamess93}
code and an uncontracted $16s12p9d4f3g$ Gaussian basis for both Li and
H - larger basis sets led to convergence difficulties in some
calculations.  Although the BFD pseudopotentials are provided with a
range of optimised basis sets, these were not used as they gave higher
total energies.  The complete active space (CAS) for the MCSCF
calculations was constructed using $20$ orbitals (spin restricted),
resulting in 132 determinants.

The VMC and DMC calculations were performed using the \textsc{CASINO}
QMC package.\cite{casino} We used the Casula scheme\cite{casula06} for
evaluating the non-local energy, which provides stable and variational
estimates of the DMC energy.  The determinants used in the trial wave
functions for the QMC calculations were taken from GAMESS MCSCF
calculations with a smaller basis ($16s12p9d$) than considered above,
and including only 5 orbitals in the active space (a CAS of 11
determinants).  In addition a Jastrow pre-factor was introduced that
includes electron-electron, electron-ion, and electron-electron-ion
terms (the form used is Eq.~(2) of Ref.~\cite{drummond04}).  The total
energy of this Jastrow/multideterminant wave function was optimised by
minimising the VMC total energy with respect to the parameters in the
Jastrow function and the multideterminant expansion coefficients using
recently developed methods.\cite{umrigar07,brown07} A final detail is
that the TN pseudopotentials are used in two forms.  These are an
``exact'' tabulated pseudopotential,\cite{trail05c} and an
accurate\cite{trail05b} Gaussian representation of the same
pseudopotential.  Since the former is the more accurate and possesses
the smaller non-local region (giving a lower computational cost), it
is used in the QMC calculations.\cite{trail05c} The Gaussian
representation is necessary for calculations involving GAMESS.

We also investigated the effect of introducing the Li CPP of Shirley
and Martin\cite{shirley93}.  The CPP attracts electrons to the core,
and therefore increases the first ionization potential of the atom.
This effect is significant in LiH because the Li core is quite
polarizable and because H has a considerably higher electronegativity
than Li and therefore tends to draw the second valence electron away
from the Li atom.  The introduction of the CPP increases the
calculated first ionization potential of the Li atom from 5.34~eV to
5.40~eV (for both TN and BFD pseudopotentials), giving a result in
good agreement with the experimental value of
5.3917~eV.\cite{lorenzen82}

\begin{figure}[t]
\includegraphics{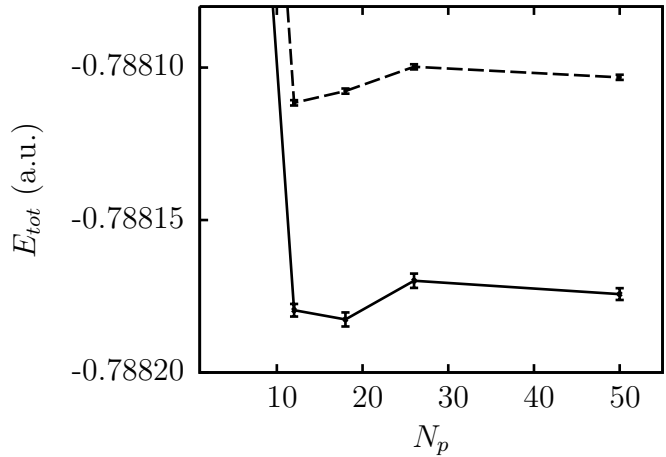}
\caption{\label{fig:1} Variation of QMC total energy estimates for LiH
with the number of integration points $N_p$.  The dashed line and
error bars denote VMC results while the solid line and error bars
denote DMC results, all with $R=1.596$~\AA\ and TN pseudopotentials
but no CPP.}
\end{figure}

We checked the convergence of the QMC calculations with respect to the
parameters of the calculations, most importantly the size of the
integration grid used for evaluating the non-local pseudopotential
energy, and the finite time step used in the DMC calculations.  In QMC
methods the non-local pseudopotential energy is evaluated via
numerical integration over the surfaces of spheres, which is
implemented in \textsc{CASINO} using well established quadrature
rules.\cite{mitas91} These rules consist of sampling on a discrete
grid of $N_p$ points, and are exact in the limit of large $N_p$.
Results for various values of $N_p$ are shown in Fig.~\ref{fig:1}.  We
found that $N_p=12$ quadrature grids which integrate the angular
momentum components of the wave function exactly up to $l=5$ were
required to give a bias in the energy smaller than 0.00001~a.u.  It is
worth pointing out that since these errors are implicitly associated
with the ionic core they may reasonably be expected to be consistent
between different systems, and will therefore tend to cancel in
estimates of the spectroscopic constants.  Unless stated otherwise,
all VMC results in this paper were obtained with $N_p=50$, which
integrates exactly up to $l=11$, for which we estimate a bias of order
0.000001~a.u.  Due to the prohibitive cost of larger grids, all of the
DMC results in this paper (unless stated otherwise) are obtained with
$N_p=12$.

\begin{figure}[t]
\includegraphics{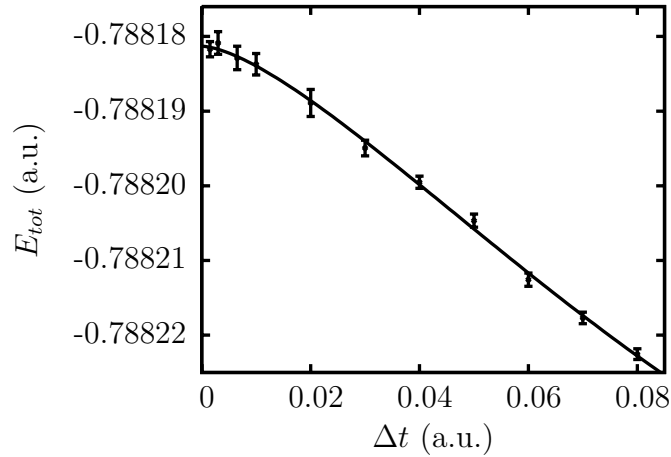}
\caption{\label{fig:2} Bias in the DMC energy as a function of time
step for LiH described by TN pseudopotentials with no CPP at a bond
length of $R=1.596$~\AA.  The solid line is the result of a
least-squares fit to the function $E_{tot}=a+b\Delta t^{1/2} + c
\Delta t + d \Delta t^{3/2} + e \Delta t^2$.  }
\end{figure}

Next we consider the convergence of DMC total energies with time step.
Figure~\ref{fig:2} shows the total energy as a function of time step
$\Delta t$ for a TN pseudopotential calculation (without a CPP) at
$R=1.596$~\AA.  The data demonstrates convergence to within a standard
error of $1.5 \times 10^{-6}$~a.u. for $\Delta t \leq 0.003$~a.u., and
we used this value for all DMC calculations unless stated otherwise.
Tests indicated that the convergence with time step is essentially the
same for other geometries and when using the BFD pseudopotentials
and/or the CPP.

\section{results}

First we compare total energies from different methods at a bond
length of $R=1.596$~\AA, using the TN pseudopotentials, see
Table~\ref{tab:2}.  Each of the computational methods is, in
principle, variational (although any bias in the QMC results may not
be), so that the lowest energy obtained is the best result.

The VMC energy is only slightly above the DMC energy, so the
Jastrow/multideterminant trial wave function is very accurate and the
error in the total energy from the approximate non-local
pseudopotential DMC scheme\cite{casula06} is small.  Furthermore, the
ground state wave function is nodeless, so there is no fixed-node
error.  It therefore seems likely that the DMC energy is only very
slightly above the exact answer.  We therefore measure the amount of
correlation retrieved by the various methods in terms of the
percentages of the DMC correlation energy obtained.  The largest
improvement in total energy beyond the HF result occurs on introducing
correlation at the MCSCF level, where we obtained $90.9$\% of the DMC
correlation energy with the 11 determinant MCSCF calculation, and
$98.2$\% with 132 determinants.  The introduction of the Jastrow
factor in the VMC calculation retrieves $99.8$\% of the correlation
energy and lowers the energy to within $10^{-4}$~a.u.\ of the DMC
energy.

The final two rows of Table~\ref{tab:2} give results obtained on
introducing the Li CPP.  The change in energy is almost the same
within VMC and DMC, and it amounts to a significant reduction of
$\sim$0.00328~a.u.  This is a measure of the influence of core-valence
correlation and core relaxation on the total energy, which amounts to
about $9\%$ of the valence correlation energy.  Note that the lowering
of the diatomic energy is about 1.6 times larger than for the isolated
Li atom, so it is clear that the CPP will significantly affect the
interatomic potential and spectroscopic constants.

\begin{table}[t]
\begin{tabular}{l|l}
Method           & $E_{\rm tot}$ (a.u.) \\ \hline \hline
HF               & $-0.7507056    $  \\
MCSCF (11 dets)  & $-0.7847907    $  \\
MCSCF (132 dets) & $-0.7874988    $  \\
VMC              & $-0.7881032(9) $  \\
DMC              & $-0.7881796(21)$  \\ \hline
VMC+CPP          & $-0.7913829(9) $  \\
DMC+CPP          & $-0.7914627(19)$  \\
\end{tabular}
\caption{ Ground state total energies from different methods for LiH
at $R=1.596$~\AA. The TN pseudopotentials are used, together with a Li
CPP for the last two rows.  The $16s12p9d4f3g$ basis set is used for
the HF and MCSCF energies given. (Using the $16s12p9d$ basis increases
the 1, 11, and 132 determinant energies by $4 \times 10^{-6}$~a.u., $2
\times 10^{-5}$~a.u., and $4 \times 10^{-6}$~a.u., respectively.) }
\label{tab:2}
\end{table}

The available experimental values for the spectroscopic constants, as
reported by Stwalley and Zemke\cite{stwalley93}, show a finite spread.
For our calculations we choose an accuracy for the solution of the
physical models (i.e., numerical precision, convergence tolerance, or
statistical error) which is less than the spread of experimental data,
corresponding to a target precision of $\sim 10^{-6}$~a.u.\ or better
for the calculated total energies.  This precision is several orders
of magnitude smaller than the energy differences between methods,
between the same method applied with different pseudopotentials, and
between the same method applied with or without the CPP.

In light of this target precision, we note that it is
possible to extend the model Hamiltonian, for example by introducing a
finite distribution of nuclear charge, or by going beyond the
Born-Oppenheimer approximation.  However, the effects of such
extensions are expected to be less than the resolution of the
calculated and experimental data, and so they have not been considered
here.

\begin{sidewaystable}
\centering
\begin{tabular}{l|l|lllllll}
Method      &
$R_e$       &$Y_{10}$      &$-Y_{20}$   &$Y_{30}$   &$Y_{01}$    &$-Y_{11}$   &$E_{\rm ZP}$ &$D_e$ \\ \hline \hline
AE HF       &
$1.606110$  &$1429.46682 $ &$20.98514 $ &$0.183708$ &$7.415712 $ &$0.183374 $ &$710.29917 $ &$11995.36673 $ \\
AE HF(rel.) &
$1.606035$  &$1429.53341 $ &$20.98503 $ &$0.182985$ &$7.416398 $ &$0.183378 $ &$710.33246 $ &$11994.38580 $ \\
AE HF(num.) &
$1.605721$  &$1429.85721 $ &$20.96747 $ &$0.176873$ &$7.419300 $ &$0.183362 $ &$710.50174 $ &$11991.64748 $ \\
HF+BFD      &
$1.598248$  &$1417.71562 $ &$20.28278 $ &$0.106127$ &$7.488848 $ &$0.177792 $ &$704.52610 $ &$12020.71124 $ \\
HF+TN       &
$1.602730$  &$1418.26816 $ &$20.57323 $ &$0.162191$ &$7.447021 $ &$0.173705 $ &$704.57600 $ &$11930.95292 $ \\ \hline
AE CI\cite{lundsgaard99}  &
$1.595275$  &$1405.6286  $ &$23.27699 $ &$0.18276 $ &$7.516788 $ &$0.217214 $ &$697.98774 $ &$20102.5526  $ \\
MCSCF+BFD   &
$1.597956$  &$1377.99761 $ &$22.72034 $ &$0.276529$ &$7.491584 $ &$0.210426 $ &$684.15908 $ &$20064.10177 $ \\
MCSCF+TN    &
$1.602560$  &$1377.69056 $ &$22.75457 $ &$0.261216$ &$7.448595 $ &$0.205308 $ &$683.87061 $ &$20006.04585 $ \\
MCSCF+CPP\cite{gadea06}  &
$1.589   $  &$    -      $ &$  -      $ &$ -      $ &$ -       $ &$ -       $ &$695.7     $ &$20350.0     $ \\ \hline
VMC+BFD     &
$1.59705(4)$&$1380.3(2)  $ &$22.4(2)  $ &$0.14(5) $ &$7.5001(3)$ &$0.2090(7)$ &$685.40(6) $ &$20196.4(1)  $ \\
VMC+TN     &
$1.59964(6)$&$1380.5(4)  $ &$22.8(3)  $ &$0.24(9) $ &$7.4758(6)$ &$0.210(1) $ &$685.4(1)  $ &$20138.3(2)  $ \\
VMC+BFD+CPP&
$1.58196(3)$&$1406.1(3)  $ &$23.8(2)  $ &$0.35(5) $ &$7.6438(3)$ &$0.2168(7)$ &$697.91(7) $ &$20484.0(1)  $ \\
VMC+TN+CPP &
$1.58491(5)$&$1405.3(4)  $ &$23.2(3)  $ &$0.14(9) $ &$7.6154(5)$ &$0.214(1) $ &$697.7(1)  $ &$20420.3(2)  $ \\
DMC+BFD+CPP&
$1.5823(1) $&$1402(1)    $ &$21.3(8)  $ &$0.3(2)  $ &$7.640(1) $ &$0.208(3) $ &$697.2(3)  $ &$20500.6(4)  $ \\
DMC+TN+CPP &
$1.5847(1) $&$1406(1)    $ &$23.8(7)  $ &$0.4(2)  $ &$7.617(1) $ &$0.216(2) $ &$697.0(2)  $ &$20437.9(4)  $ \\ \hline
CD Model\cite{coxon04} &
$1.595601  $&$1405.51646 $ &$23.18135 $ &$0.173478$ &$7.513712 $ &$0.216316 $ &$697.95379 $ &$20286.0000  $ \\
Exp.\cite{stwalley93} &
$1.595584  $&$1405.50936 $ &$23.17938 $ &$0.176365$ &$7.513751 $ &$0.216461 $ &$697.95    $ &$20287.7(3)  $ \\ \hline \hline
\end{tabular}
\caption{Estimates of $\{Y_{ij}\}$, $E_{\rm ZP}$, $R_e$, and $D_e$
from various calculational methods, the CD model potential and
experiment. The $\{Y_{ij}\}$, $E_{\rm ZP}$, and $D_e$, are in
cm$^{-1}$ and $R_e$ is in \AA.  [1.0~cm$^{-1}$ =
4.556335~$\times$~10$^{-6}E_h$\ = 1.239842 $\times$ 10$^{-4}$~eV.]}
\label{tab:3}
\end{sidewaystable}

Table~\ref{tab:3} gives the spectroscopic constants obtained from
various calculations, the CD model potential\cite{coxon04}, and
experiment.\cite{stwalley93} In each case the isolated atom energies
used to obtain $U_{\infty}$ in Eq.~(\ref{eq:4}) were obtained from
either a Gaussian basis set and the GAMESS code, a numerical
integration and the ATSP2K code,\cite{atsp2k} or the best available
values in the literature (all calculated using the appropriate
Hamiltonian).  $E_{\rm ZP}$ and $R_e$ were obtained from the
$\{Y_{ij}\}$, but $D_e$ was obtained from the fitted potential.  Both
of the pseudo-atoms possess only a single electron, hence only HF
calculations were required.  All-electron atomic HF results were
calculated numerically or with a Gaussian basis, as appropriate, and
the post-HF AE atomic energies are from the same papers as the
molecular data.  No AE QMC calculations were performed.

The data in Table~\ref{tab:3} allows a direct comparison of estimates
of the vibrational and rotational properties, the equilibrium
geometry, and dissociation energies.  In order to be consistent with
the experimental results the `harmonic equilibrium separation' of
Eq.~(\ref{eq:3}) is reported rather than interatomic separation with
lowest total energy, but the difference between these two quantities
is less than $10^{-5}$~\AA.

The experimental results taken as benchmark data are found in the
bottom two lines of the table.  The ``CD Model'' values are extracted
from the CD model potential\cite{coxon04} using exactly the same
procedure as for the \emph{ab initio} results.  The ``Experimental''
values are those recommended in the literature for low level
excitations, which are the most appropriate for our results, i.e., the
results obtained by Maki \emph{et al.} and reviewed by
Stwalley\cite{stwalley93}.

Hartree-Fock results for the TN and BFD pseudopotentials are reported
in Table~\ref{tab:3}.  Also shown are the AE results obtained using
the same basis set as for the pseudopotential results (from GAMESS),
using the same Gaussian basis and scalar relativistic corrections
(from the Douglas, Kroll, and Hess Hamiltonian and GAMESS), and using
accurate numerical integration (from the 2DHF\cite{2dhf} package)
without relativistic corrections.  This set of results allows a
separation of the errors due to the pseudopotentials generally, due to
the differences between the TN and BFD pseudopotentials, due to the
use of a finite basis set, and due to relativistic effects.  A
separation of errors within HF theory is useful since any
approximation that fails at this level is unlikely to succeed at a
higher level of theory, and because the uncorrelated energies are
certainly obtained very accurately.

The HF data gives poor approximations to the experimental results, and
the calculations are precise enough to draw some conclusions about the
sources of these errors.  The difference between the finite basis set
and numerical AE HF results is approximately equal to the target
accuracy, and is small compared to the difference between the AE HF
and experimental or pseudopotential results.  From this it seems
reasonable to conclude that basis set errors are not significant at
the available resolution, especially when we bear in mind that the
Gaussian basis used was constructed and tested for use with
pseudopotentials. Similarly, the difference between the relativistic
and non-relativistic AE HF results is less than the target precision,
and hence we conclude that relativistic effects are not significant at
the available resolution.

The HF results for the TN and BFD pseudopotentials are very similar,
so it seems reasonable to ascribe most of the consistent difference
between the AE HF and pseudopotential HF results to core relaxation.
A similar level of consistency occurs between results for both
pseudopotentials when we include correlation at different levels,
suggesting that we may ascribe most of the consistent difference
between these results and the experimental results to the combination
of core relaxation and core-valence correlation, and so distinguish
between the two effects.  For example, $Y_{10}$ shows an increase of
$11.9(6)$~cm$^{-1}$ due to core relaxation, and an increase of
$13.2(4)$~cm$^{-1}$ due to core-valence correlation, to be compared
with a reduction of $-37.6(4)$~cm$^{-1}$ due to valence correlation.
(Where valence correlation is defined via the VMC results, the quoted
values are the mean of the values for the two pseudopotentials, and
the bracket gives the absolute difference between them.)

Table~\ref{tab:3} also contains results obtained from the AE full CI
total energies of Ref.~\cite{lundsgaard99}. These data were used
because they provide the most accurate \emph{ab initio} estimates of
the spectroscopic constants that we are aware of.  These results are
in good agreement with experiment, except for $D_e$ which is
underestimated.  The MCSCF results with the two pseudopotentials agree
well with each other, but differ significantly from the AE CI
results.

Similarly, the VMC results for the TN and BFD pseudopotentials
(without the Li CPP) agree well with one another and are consistent
with the analogous MCSCF results.  The only clear improvement in the
VMC results over the MCSCF ones is in the well depth, $D_e$, which
might be expected from the more complete description of correlation
effects provided by the VMC calculations.

For each level of correlated calculation, the values of $D_e$ and
$R_e$ calculated using the TN and BFD pseudopotentials differ by more
than the target accuracy, and therefore we can distinguish differences
in their performance.  However, the results obtained with the two
pseudopotentials are much closer to each other than they are to
experiment, so it is clear that neither pseudopotential gives highly
accurate results.

We note that the TN pseudopotential consistently gives slightly larger
values of $R_e$ than the BFD one, by $+0.00448$~\AA\ in HF theory,
$+0.00460$~\AA\ in MCSCF and $+0.00295$~\AA\ for VMC (without the
CPP).
Burkatzki \textit{et al.}\cite{burkatzki07} studied a number of
molecules, including LiH, and found the LiH bond length with the TN
pseudopotentials to be $0.0361$~\AA\ larger than for the BFD ones.
This difference is an order of magnitude larger than we have found,
and we believe their bond-length difference must be biased.
Similarly, the harmonic vibrational frequencies ($Y_{10}$) and well
depth ($D_e$) calculated for TN and BFD by Burkatzki \textit{et al.}
differ by $-18.1$~cm$^{-1}$ and $+790$~cm$^{-1}$, respectively,
whereas our VMC results (without the CPP) differ by $+0.2$~cm$^{-1}$
and $-0.166$~cm$^{-1}$, respectively.
Again we suggest that the harmonic vibrational frequencies and well
depths calculated by Burkatzki \textit{et al.} are biased in some way.

% conversion done with 1hart   = 627.50956  kcal/mol
%                      1cm{-1} =   4.556335 10^{-6} hart

\begin{figure}[t]
\includegraphics{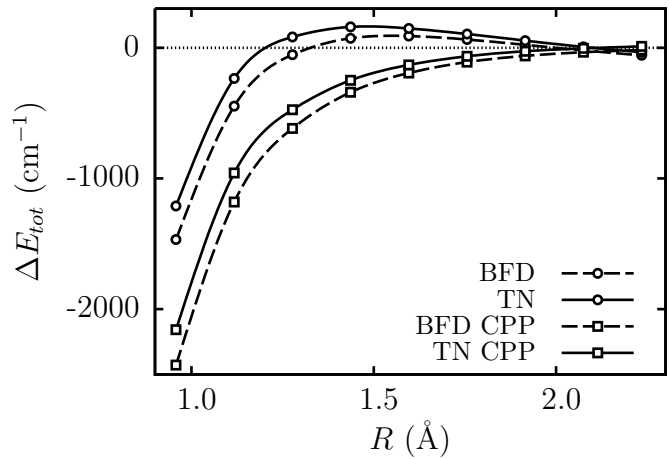}
\caption{\label{fig:3} Difference between the VMC total energies,
$\Delta E_{tot}$, and those provided by the CD model potential of
\cite{coxon04}.  Energy differences are shown for calculations
performed with both TN (solid lines) and BFD (dashed lines)
pseudopotentials, and both with (squares) and without (circles) the Li
CPP potential.  An offset is added such that $\Delta E_{tot}
\rightarrow 0$ as $R \rightarrow \infty$.
}
\end{figure}

Introducing the Li CPP\cite{shirley93} into the QMC calculations
improves $E_{\rm ZP}$ and $Y_{10}$ significantly, but makes $D_e$ too
large and $R_e$ too small.  These ``biases'' in the calculated values
are consistent with those found in previous MCSCF
calculations\cite{gadea06} using a different pseudopotential and CPP
to those used here.
It is therefore reasonable to attribute these biases in our results to
a general deficiency of the pseudopotential+CPP model for the ionic
cores.

Figure~\ref{fig:3} shows these biases as the difference between the
VMC total energies and the CD model potential\cite{coxon04}, with an
energy offset added such that all curves approach zero for increasing
$R$.  All results show an underestimate of the total energy with
decreasing $R$ of rapidly increasing magnitude, which we ascribe to
the overlap of the pseudopotential core regions.  For the calculations
without the CPP an overestimate occurs near equilibrium, whereas when
the CPP is included this becomes an underestimate of larger magnitude,
and for all $R$.  This general trend occurs for both pseudopotentials,
such that at the equilibrium separation the BFD result is better than
the TN result for no CPP, but the reverse is true when the CPP is
introduced.  The same trend also occurs for the equilibrium separation
itself.  Note that the improved CPP estimates for $Y_{10}$, $Y_{20}$,
$Y_{11}$, and $E_{\rm ZP}$ correspond to superior 2$^{nd}$ and higher
order derivatives of the potential at $R_e$, so are more subtle in
Fig.~\ref{fig:3}.

For all geometries considered, the change in total energy upon
introduction of the CPP is very similar for both pseudopotentials.
This is consistent with the suggestion of Shirley and
Martin\cite{shirley93} that their CPPs are valid for use with
different types of HF pseudopotential.

For many systems the computational cost of evaluating the integral
required for the pseudopotential energy is very significant, and may
even be dominant for DMC calculations in which the number of points in
the integration grid, $N_p$, is large.  The computational cost of the
non-local integration depends on the size of the region inside the
non-local radius, which is the distance beyond which the angular
momentum components of the pseudopotential differ by less than some
parameter $\epsilon$.  The non-local radii of the \textit{tabulated}
TN pseudopotentials are consistently smaller than those of the BFD
pseudopotentials. For $\epsilon=10^{-5}$~a.u. and Li, the non-local
radii are $1.44$~\AA\ (TN) and $1.66$~\AA\ (BFD). For the DMC
calculations considered here this resulted in a $14\%$ smaller
computational cost when using the TN pseudopotentials. The advantage
is smaller for VMC calculations.  The cutoff radii of the TN
pseudopotentials also depend less sensitively on the value of
$\epsilon$. For example, with a tolerance of $10^{-4}$~a.u.\ the Li
core radii are $1.40$~\AA\ (TN) and $1.52$~\AA\ (BFD) ($9\%$ greater),
while for a tolerance of $10^{-8}$~a.u.\ the two radii are $1.54$~\AA\
and $2.06$~\AA, respectively (a $34\%$ increase).  We emphasise that
the tabulated TN pseudopotentials are intended for use in QMC
calculations,\cite{trail05c} while the Gaussian parameterisations are
intended for use in quantum chemistry codes which require such a
representation.  For the latter, the non-local radius is larger,
taking the value $1.69$~\AA\ for $\epsilon=10^{-5}$~a.u.

\section{Conclusions}
Pseudopotentials are often used in \emph{ab initio} calculations with
very little account taken of the error introduced.  In many cases,
particularly for explicitly correlated calculations, the errors from
the pseudopotential and the approximate method used to calculate the
energies are not distinguished.  We have chosen a system which is
sufficiently small that we can solve it to very high accuracy using a
variety of methods and for which a large amount of accurate
experimental data is available.  Anharmonic effects are very important
for the LiH ground state and the prediction of its spectroscopic
constants is a severe test of the theory.  It is also worth mentioning
that in the process of extracting the spectroscopic constants from the
\emph{ab initio} total energies an interatomic potential is generated
with controlled accuracy.

We have compared pseudopotentials constructed from uncorrelated atomic
calculations using different approaches, the ``shape-consistent'' or
norm-conserving pseudopotentials of TN, and the ``energy-consistent''
pseudopotentials of BFD.  The differences between the results obtained
with the TN and BFD pseudopotentials are small compared with the
differences from experiment.  This conclusion holds whether or not the
CPP is included.  Introducing the CPP substantially improves the
zero-point energy $E_{\rm ZP}$, but the harmonic equilibrium
separation $R_e$, and well-depth $D_e$, are poorer.

The changes in the QMC estimates of $R_e$, $w_e$, and $D_e$ upon
introducing the CPP are consistent with those found by previous
authors using quantum chemistry methods, and their analysis of the
source deficiencies of the CPP.\cite{othercpp} However, the accurate
description of the electron-electron cusp in the QMC calculations
significantly improves the estimates of $w_e$ and $D_e$ over those
found using quantum chemistry methods.

It appears that using a HF-based pseudopotential which does not
contain correlation effects and adding the CPP which approximately
describes core relaxation and core-valence correlation effects
provides a reasonable model of LiH, but one which still suffers from
significant errors.  Whether these errors can be reduced by using a Li
pseudopotential constructed using data from correlated calculations
and/or by modifying the Li CPP is currently unknown.

\begin{acknowledgements}
This work was supported by the Engineering and
Physical Sciences Research Council (EPSRC) of the United Kingdom, and
computing resources were provided by the HPCx Consortium.
\end{acknowledgements}

%\newpage
%\begin{figure}[t]
%\includegraphics{./fig1.ps}
%\caption{\label{fig:1} 
%Fig. 1: Variation of QMC total energy estimates for LiH
%with the number of integration points $N_p$.  The dashed line and
%error bars denote VMC results while the solid line and error bars
%denote DMC results, all with $R=1.596$~\AA\ and TN pseudopotentials
%but no CPP.}
%\end{figure}

%\newpage
%\begin{figure}[t]
%\includegraphics{./fig2.ps}
%\caption{\label{fig:2} 
%Fig. 2: Bias in the DMC energy as a function of time
%step for LiH described by TN pseudopotentials with no CPP at a bond
%length of $R=1.596$~\AA.  The solid line is the result of a
%least-squares fit to the function $E_{tot}=a+b\Delta t^{1/2} + c
%\Delta t + d \Delta t^{3/2} + e \Delta t^2$.}
%\end{figure}

%\newpage
%\begin{figure}[t]
%\includegraphics{./fig3.ps}
%\caption{\label{fig:3} 
%Fig. 3: Difference between the VMC total energies,
%$\Delta E_{tot}$, and those provided by the CD model potential of
%\cite{coxon04}.  Energy differences are shown for calculations
%performed with both TN (solid lines) and BFD (dashed lines)
%pseudopotentials, and both with (squares) and without (circles) the Li
%CPP potential.  An offset is added such that $\Delta E_{tot}
%\rightarrow 0$ as $R \rightarrow \infty$.}
%\end{figure}

\end{document}